# Approximate method of free energy calculation for spin system with arbitrary connection matrix

Boris Kryzhanovsky and Leonid Litinskii

Center for Optical Neural Technologies, Scientific Research Institute for System Analysis of RAS, Vavilov str., 44/2, off. 216, Moscow, Russia, 119333

kryzhanov@mail.ru, litin@mail.ru

**Abstract**. The proposed method of the free energy calculation is based on the approximation of the energy distribution in the microcanonical ensemble by the Gaussian distribution. We hope that our approach will be effective for the systems with long-range interaction, where large coordination number q ensures the correctness of the central limit theorem application. However, the method provides good results also for systems with short-range interaction when the number q is not so large.

## 1. Introduction

We examine a system of $N$ spins $s_i = \{\pm 1\}$, whose state is determined by the configuration vector $\mathbf{s} = (s_1,...,s_N)$ and the energy $E(s) = -\frac{1}{2}\sum_{i=1}^{N} s_i h_i$. Here $h_i = \sum_{i=1}^{N} J_{ij} s_j$ is the local field acting on the $i$-s spin. We suppose that $\mathbf{J} = (J_{ij})$ is a symmetric matrix; the diagonal elements of this matrix are equal to zero ($J_{ii} = 0$) and in average its rows contain a large number of nonzero elements $q \gg 1$. The inequality $q \gg 1$ means that correlation between $s_i h_i$ and $s_j h_j$ is negligible small (it behaves as $q^{-1}$).

Let us describe our method of calculation of the partition function $Z_N = \sum_{\mathbf{s}} \exp[-\beta E(\mathbf{s})]$, where $\beta$ is the inverse temperature. Let $\mathbf{s}_0$ be the ground state. The class of all states $\mathbf{s}$ differing from $\mathbf{s}_0$ by $n$ opposite spin signs we define as $\Omega_n$:

$$\Omega_n = \{\mathbf{s}: (\mathbf{s}, \mathbf{s}_0) = N - 2n\}, n = 0, 1, ... N.$$

In the general case we do not know the form of the energy distribution for all $\binom{N}{n}$ states of the class $\Omega_n$. However, the exact expressions of the mean value $E_n$ and dispersion $\sigma_n^2$ can be calculated (see [1], [2]):

$$E_n = E(\mathbf{s}_0) \frac{(N-2n)^2 - N}{N(N-1)},$$

$$\sigma_n^2 = \frac{2n(N-n)}{N(N-1)(N-2)(N-3)} \left[ A \left( \sum_{i,j=1}^{N} J_{ij}^2 - \frac{4E(\mathbf{s}_0)^2}{N(N-1)} \right) + B \left( \|\mathbf{Js}_0\|^2 - \frac{4E(\mathbf{s}_0)^2}{N} \right) \right],$$

$$A = 4(n-1)(N-n-1), \quad B = 2\left[ (N-2n)^2 - N + 2 \right].$$

Further on, it can be shown that for large values of $n$ and $N$ the distribution of the energies for the class $\Omega_n$ can be approximated reasonable well by the Gaussian distribution with the parameters $E_n$ and $\sigma_n^2$ [1].

We verified this assertion for some types of the matrices $\mathbf{J}$, the dimensionalities $N \sim 10^3 - 10^4$ and different values of $n \gg 1$. At least if $E \in [E_n \pm 3\sigma_n]$, the experimental density of the distribution (for large number of random configurations $K \geq 10^6$) coincides perfectly with the density of the Gaussian distribution (see figure 1). Outside of this interval the densities differ, however for our purposes these distinctions on the tails are not pivotal.

Then the summation over the class $\Omega_n$ can be replaced by integration over the Gaussian measure, and the expression for the partition function takes the form

$$Z_N = \sum_{n=0}^{N} \sum_{\Omega_n} \exp[-\beta E(\mathbf{s})] \approx \sum_{n=0}^{N} \binom{N}{n} \int_{E_{\min}}^{E_{\max}} \exp(-\beta E) \exp\left(-\frac{(E-E_n)^2}{2\sigma_n^2}\right) \frac{dE}{\sqrt{2\pi}\sigma_n}, \quad (1)$$

where the minimal and maximal values of the energy $E(\mathbf{s})$ are taken as the inferior and superior limits of integration. This technique is essential for our approach. Now the calculation of the partition function reduces to the calculation of the integral with the aid of the saddle-point method.

## 2. The Main Equation

Let us define asymptotic expressions for $E_n$ and $\sigma_n^2$, normalized to $N$. They are $E_x = \lim_{N\to\infty} E_n/N$ and $\sigma_x^2 = \lim_{N\to\infty} \sigma_n^2/N$, where $x = n/N$. Then we have:

$$E_x = E_0(1-2x)^2, \quad \text{where } E_0 = E(\mathbf{s}_0)/N,$$
$$\sigma_x^2 = \left(N^{-1}\sum_{i\neq j}^{N} J_{ij}^2\right) 2x(1-x)\left[4x(1-x)(1-\varepsilon_0^2) + 2N(d_0 - \varepsilon_0^2)(1-2x)^2\right], \quad (2)$$
$$\text{and} \quad \varepsilon_0^2 = 4E_0^2\left(\sum_{i\neq j}^{N} J_{ij}^2\right)^{-1}, \quad d_0 = \|\mathbf{Js}_0\|^2 \left(N\sum_{i\neq j}^{N} J_{ij}^2\right)^{-1}.$$

Using the Stirling asymptotic and replacing the summation over $n$ with the integration over the variable $x$, we obtain:

$$Z_N \sim \sqrt{\tfrac{N}{2\pi}} \int_0^1 e^{-Nf(x)} dx,$$
$$f(x) = L(x) + \beta E_x - \tfrac{1}{2}\beta^2\sigma_x^2 - N^{-1}\ln\Phi(x), \quad (3)$$

where $L(x) = x\ln x + (1-x)\ln(1-x)$. Finally, we set $E'_{\max} = E_{\max}/N$, and then

$$\Phi(x) = \frac{1}{\sqrt{2\pi}} \int_{A_x}^{B_x} e^{-\frac{1}{2}t^2} dt, \quad B_x = \sqrt{N}\left(\beta\sigma_x + \frac{E'_{\max}-E_x}{\sigma_x}\right), \quad A_x = \sqrt{N}\left(\beta\sigma_x + \frac{E_0 - E_x}{\sigma_x}\right).$$

Let us calculate the limits of integration $B_x$ and $A_x$ as well as the function $\Phi(x)$. Since $E'_{\max} - E_x > 0$, the numerator in the expression for $B_x$ is always positive, and $B_x \to +\infty$ when $N \to \infty$. The numerator in the expression for $A_x$ can be both negative and positive. In the first case $A_x \to -\infty$ when $N \to \infty$, and then $\Phi(x) \to 1$. But if the numerator of $A_x$ is positive, the value of $A_x \to +\infty$ and $\Phi(x) \sim \exp\left(-\tfrac{1}{2}A_x^2\right)$. Substituting these expressions for $\Phi(x)$ in (3), we finally obtain:

$$f(x) = \begin{cases} f^{(1)}(x) = L(x) + \beta E_x - \tfrac{1}{2}\beta^2\sigma_x^2, & \text{when } \beta\sigma_x^2 + E_0 - E_x < 0 \\ f^{(2)}(x) = L(x) + \beta E_0 - \dfrac{1}{2}\left(\dfrac{E_0 - E_x}{\sigma_x}\right)^2, & \text{when } \beta\sigma_x^2 + E_0 - E_x > 0 \end{cases} \quad (4)$$

The free energy per site is equal to $f(\beta) = \beta^{-1} \min_x f(x)$.

## 3. The Ising Model on a Hypercube

Our approach is theoretically based for $q \gg 1$. Nevertheless, rather good results can also be obtained for comparatively small values of $q$.

For $D$-dimensional Ising model in each row of the matrix $\mathbf{J}$ there are only $q = 2D$ nonzero elements, and they all are equal to $J$. Here the ground state is the configuration $\mathbf{s}_0 = (1,1,...,1)$. It is easy to obtain that $E_0 = -\frac{1}{2}qJ$, $\varepsilon_0^2 = q/N \to 0$, $d_0 - \varepsilon_0^2 = 0$, and $\sigma_x^2 = 8qJ^2 x^2 (1-x)^2$. The function $f(x)$ (4) is a symmetrical one with respect to the middle of the interval $x \in [0,1]$, and this is the reason why we present the expression for $f(x)$ only inside the interval $0 \le x \le \frac{1}{2}$. Let us introduce dimensionless parameter $\bar{\beta} = \beta J$. Then we obtain for (4):

$$f(x) = \begin{cases} f^{(1)}(x) = L(x) - \frac{1}{2} q \bar{\beta} \left[ (1-2x)^2 + 8 \bar{\beta} x^2 (1-x)^2 \right], & \text{for} \quad 0 \le x < x_{\bar{\beta}} \\ f^{(2)}(x) = L(x) - \frac{1}{2} q (\bar{\beta} - \frac{1}{2}), & \text{for} \quad x_{\bar{\beta}} \le x \le \frac{1}{2} \end{cases} \quad (5)$$

Here $x_{\bar{\beta}} = 1/2$, when $\bar{\beta} < 1$, and $x_{\bar{\beta}} = \left(1 - \sqrt{1 - 1/\bar{\beta}}\right)/2$, when $\bar{\beta} > 1$.

It is easy to solve the equation $df/dx = 0$ and to examine the properties of the solution (see [7]). At first let us present the final result for $q > 16/3$. When $\bar{\beta}$ is not very large, the function $f^{(1)}(x)$ has only one minimum at the point $x_0 = 1/2$ (figure 2a). When $\bar{\beta}$ becomes larger than the critical value

$$\bar{\beta}_c = \frac{1 - \sqrt{1 - 4/q}}{2}, \quad (6)$$

the point $x_0 = 1/2$ becomes the maximum point of the function $f^{(1)}(x)$, and near it appears the minimum point of this function. We denote the new minimum point $x_1(\bar{\beta})$. The phase transition of the second kind takes place when $\bar{\beta} = \bar{\beta}_c$. When $\bar{\beta}$ increases the minimum point $x_1(\bar{\beta})$ moves farther away from the point $x_0$ (figure 2b). When $\bar{\beta} \to \infty$ the minimum point $x_1(\bar{\beta}) \to 0$, and the state of the system tends to the ground state $\mathbf{s}_0 = (1,1,...,1)$ (figure 2c). This corresponds to the physical description of the spin system: it passes into the ground state when temperature tends to zero.

For 3D-Ising model ($q = 6$) from Eq.(6) we obtain $\bar{\beta}_c \approx 0.2113$, which is close to the value 0.2224 obtained in [4] with the aid of computer simulations. For 4D-Ising model ($q = 8$) we have $\bar{\beta}_c \approx 0.146$, and the difference between this value and the estimate 0.149 obtained in [5] is even less. In general, when $q$ is large enough, from Eq.(6) we obtain $\bar{\beta}_c \approx 1/q$. This is in agreement with the supposition that for large dimensionalities $D$ the Ising model shows indices typical for the mean field model [5], [6].

Now let us describe the results for $q < 16/3$. For 2D-Ising model ($q = 4$) there are some notable distinctions in the scenario described above. (Because of the length of this paper it is impossible to present them in this paper.) Here no analytic expression for the critical value $\bar{\beta}_c$ can be obtained; $\bar{\beta}_c \approx 0.3912$ is a solution of a transcendental equation. This value differs from the known exact solution 0.4407 [3]. For 1D-Ising model ($q = 2$) our approach does not work at all: the state of the spin system does not tend to the ground state when $\bar{\beta} \to \infty$. It is not surprisingly, since our approach can be justified only when $q \gg 1$.

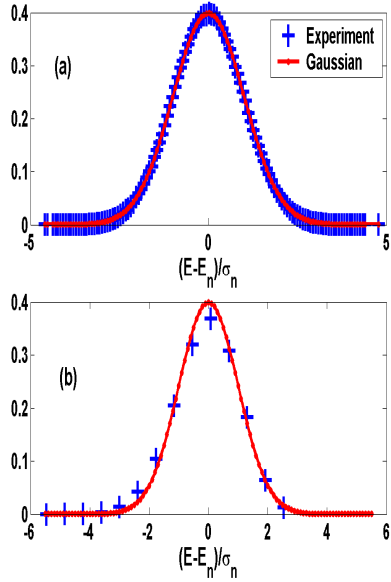
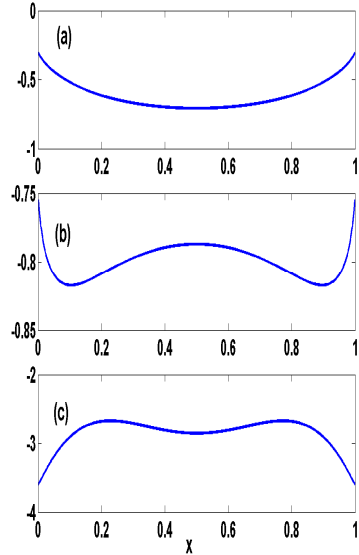

**Figure 1.** Experimental and Gaussian densities: (a) 3D-Ising, $N=10^3, n=500, K=10^6$; (b) 2D-Ising, $N=10^4, n=15, K=32\cdot 10^6$.

**Figure 2.** The function $f(x)$ (5) for 3D-Ising: (a) $\bar\beta<\bar\beta_c$, (b) $\bar\beta>\bar\beta_c$, and (c) $\bar\beta>>\bar\beta_c$.

When in our model we introduce an external nonuniform magnetic field, the expressions for $E_n$ and $\sigma_n^2$ as functions of $\mathbf{H}=(H_1,...,H_N)$ have the form:

$$E_n(\mathbf{H}) = E_n - \left(1 - \frac{2n}{N}\right)\cdot(\mathbf{s}_0,\mathbf{H}),$$

$$\sigma_n^2(\mathbf{H}) = \sigma_n^2 + \frac{4n(N-n)}{N(N-1)}\left[\|\mathbf{H}\|^2 - \frac{(\mathbf{s}_0,\mathbf{H})^2}{N} + 4\frac{N-2n}{N-2}\left(\frac{(\mathbf{Js}_0,\mathbf{H})}{2} + \frac{E(\mathbf{s}_0)\cdot(\mathbf{s}_0,\mathbf{H})}{N}\right)\right].$$

These expressions are significantly simpler when $\mathbf{H}$ is collinear to $\mathbf{s}_0$: $\mathbf{H}=H\cdot\mathbf{s}_0$. Then it is easy to show, that $E_n(\mathbf{H})=E_n-(1-2n/N)NH$ and $\sigma_n^2(\mathbf{H})=\sigma_n^2$. In this case an additional term $-\beta H(1-2x)$ appears in Eq.(3) for $f(x)$. Near the critical point $\bar\beta_c$ (6) expressions for spontaneous magnetization $m_0$, magnetic susceptibility $\chi$ and specific heat can be obtained [7]. The corresponding critical indices are the classical ones: if $t=1-\bar\beta_c/\bar\beta<<1$, then $m_0\sim t^{1/2}$, $\chi(t)\sim t^{-1}$, and a specific heat discontinuity corresponds to $\alpha=0$.

The analysis of this section can be applied to any other model when in each row of the matrix $\mathbf{J}$ there are only $q$ nonzero, equal to each other elements. Then the configuration $\mathbf{s}_0=(1,1,...,1)$ is the ground state again, and Eq.(5) is still true. In particular, it works for Ising model on the Bethe lattice [3], when each spin interacts with $q$ nearest neighbors. Equation (6) is in qualitative accord with the exact expressions obtained for the Bethe lattice. Both results coincide, when $q\to\infty$.

When for any $n$ we have $\sigma_n\to 0$, the expression (1) takes the form well-known in the mean field theory [3]: $Z_N=\sum_{n=0}^N \binom{N}{n}e^{-\beta E_n}$. From Eq.(2) we can see that the dispersion $\sigma_x^2$ vanishes only if all non-diagonal matrix elements are equal to each other: $q = N-1$, $J_{ij} \equiv J$ ($i \neq j$). In this case $\mathbf{s}_0 = (1,1,...,1)$, $d_0 - \varepsilon_0^2 = 0$, and $\varepsilon_0^2 = 1$. This simplifies all the calculations, and we easily obtain the classical Bragg-Williams equation [3]. For all other matrices the dispersion $\sigma_x^2 \neq 0$, and our approach allows one to obtain more reasonable estimates than the mean field model.

The work is supported by the Russian Basic Research Foundation (grants 12-07-00295 and 13-01-00504).